\documentclass[runningheads]{llncs}
\usepackage[T1]{fontenc}
\usepackage{graphicx}
\DeclareMathAlphabet{\mathdutchcal}{U}{dutchcal}{m}{n}
\usepackage[ruled,linesnumbered]{algorithm2e}

\usepackage{caption}
\usepackage{float}
\usepackage{subcaption}
\usepackage{amsfonts}
\usepackage{amsmath}
\usepackage{amssymb}
\usepackage{amsthm}
\usepackage{amsbsy}
\usepackage{enumerate} \usepackage{multicol}
\usepackage{mathrsfs} \usepackage[all,cmtip]{xy}

\usepackage{xpatch}
\usepackage{wrapfig}

\title{Tangent phylogenetic PCA}

\author{Morten Akhøj \inst{1,2} \and
Xavier Pennec \inst{1} \and
Stefan Sommer\inst{2}}
\authorrunning{M. Akhøj et al.}
\institute{Université Côte d'Azur and Inria, Sophia Antipolis, France \\ \email{\{morten.pedersen, xavier.pennec\}@inria.fr} \and
Department of Computer Science, University of Copenhagen, Denmark
\\
\email{sommer@di.ku.dk}}

\begin{document}

\maketitle

    \begin{abstract} 
        Phylogenetic PCA (p-PCA) is a version of PCA for observations that are leaf nodes of a phylogenetic tree. P-PCA accounts for the fact that such observations are not independent, due to shared evolutionary history. The method works on Euclidean data, but in evolutionary biology there is a need for applying it to data on manifolds, particularly shapes.
        We provide a generalization of p-PCA to data lying on Riemannian manifolds, called \textit{Tangent p-PCA}. Tangent p-PCA thus makes it possible to perform dimension reduction on a data set of shapes, taking into account both the non-linear structure of the shape space as well as phylogenetic covariance. We show simulation results on the sphere, demonstrating well-behaved error distributions and fast convergence of estimators. Furthermore, we apply the method to a data set of mammal jaws, represented as points on a landmark manifold equipped with the LDDMM metric.
    \end{abstract}

\section{Introduction} 

Phylogenetic PCA (\cite{revell2009size}, \cite{polly2013phylogenetic}) is a method for doing Principal Component Analysis (\cite{pearson1901liii}, \cite{hastie2009elements}) on data that is assumed to be dependent in a particular way. Namely, the observations $x_1, \dots, x_N \in \mathbb{R}^d$ are assumed to be the leaf nodes of a phylogenetic tree, i.e. a rooted, bifurcating tree, as exemplified on figure 1a. According to the Brownian motion model of evolution (\cite{cavalli1967phylogenetic}, \cite{felsenstein1973maximum}), the branches of this tree represents Brownian motions: the value of a node $n \in \mathbb{R}^d$ is the endpoint of a Brownian motion starting at the parent node $m \in \mathbb{R}^d$. The branch length between the nodes equals the duration of the Brownian motion. The values of two leaf nodes $x_i, x_j$ in such a tree are not independent; the more recent their latest common ancestor is, the closer they are likely to be. Phylogenetic PCA is designed to filter out this phylogenetic covariance among observations; it instead describes the variability that is \textit{not} due to the evolutionary covariance.  

Phylogenetic PCA (p-PCA) is defined for variables taking values in a Euclidean space. We generalize p-PCA to the setting where the nodes of the tree can take values in a general Riemannian manifold, not only $\mathbb{R}^d$. In studies of morphological evolution, p-PCA has so far been applied to landmark shapes which are treated as elements of a Euclidean space \cite{polly2013phylogenetic}. From a mathematical point of view, this is a crude approximation which ignores the non-linear structure of shapes (adding or subtracting two shapes does not necessarily result in a meaningful shape). In this work, we treat a shape as a point on a Riemannian manifold, e.g. Kendall's shape space \cite{kendall1984shape} or an LDDMM shape space (\cite{younes2010shapes} and \cite{miller2002metrics}). \textit{Tangent phylogenetic PCA} thus enables doing dimension reduction for shapes properly treated as elements of a manifold, while accounting for the non-independence caused by the underlying tree structure.

\begin{figure}
  \centering
  \subfloat[a][]{\includegraphics[scale=0.33]{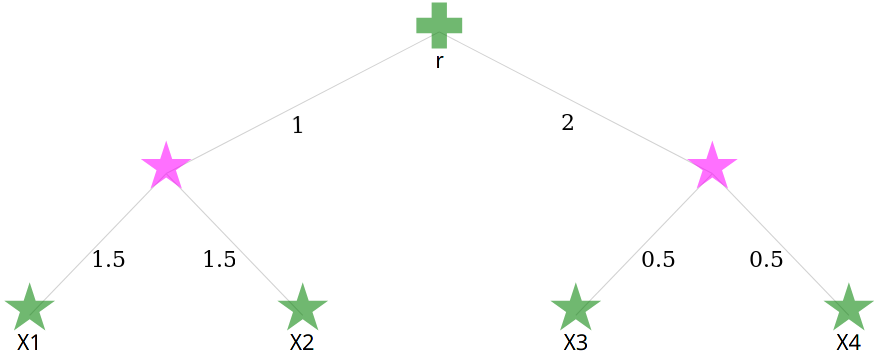} \label{fig:a}} \\
  \subfloat[b][]{\includegraphics[scale=0.4]{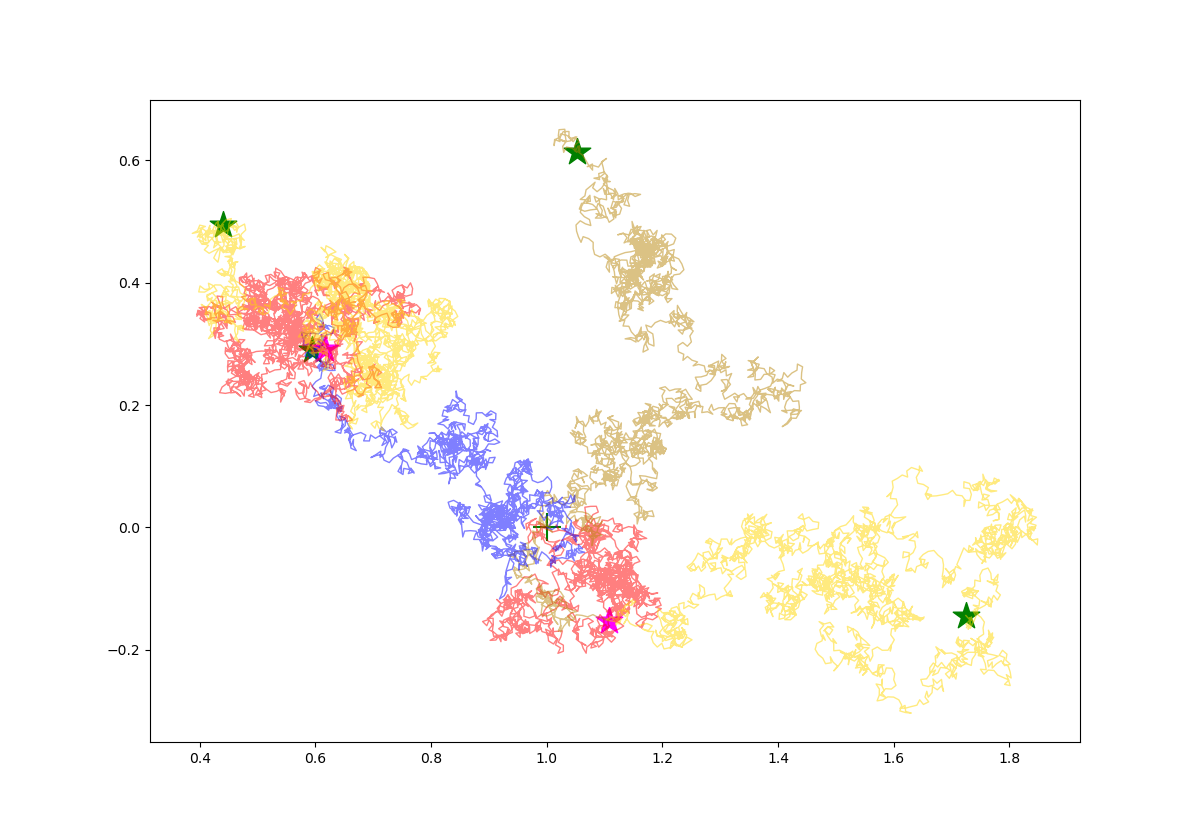} \label{fig:b}}
  \caption{\textit{a}). An example of a phylogenetic tree. \textit{r} is the root node and $x_1, \dots, x_4$ are the leaf nodes. The edge lengths are written next to each edge. \textit{b}). A realization of the tree in figure a), generated from Brownian motions in $\mathbb{R}^2$. The green '+' at $(1,0)$ is the root node. The two pink stars are the inner nodes. The 4 green stars are the leaf nodes. There are 6 Brownian motions, corresponding to the 6 edges in figure a). If two edges in figure a) share a node, the corresponding Brownian motions has different colours.} \label{fig:AB}
\end{figure}
Brownian motions are well-defined on Riemannian manifolds, so the statistical model of Brownian motions combined to form a bifurcating tree is directly transferable to the manifold setting. Our extension of p-PCA, \textit{tangent p-PCA}, works by iteratively using linear approximations of the manifold. In each linear approximation, the Euclidean formulas for p-PCA are used. This allows for retaining the interpretation from the Euclidean setting, while making sure that the output respects the non-linear structure of e.g. shapes.

    \subsection{Overview of the paper}
    
    In section \ref{sect_p_PCA}, we describe phylogenetic PCA in the Euclidean setting. In section \ref{sect_riemannian_geometry} we describe the necessary elements of Riemannian geometry, in particular geodesics, the geodesic distance and the Riemannian Brownian motion. The section also includes basic elements of statistics on a Riemannian manifold. In section \ref{sect_tangent_p_PCA} we describe tangent phylogenetic PCA. In section \ref{sect_applications} we illustrate the method on simulated data on the sphere, and apply it to a data set of mammal jaws, represented by landmarks in the LDDMM shape space.
    
\section{Phylogenetic PCA}\label{sect_p_PCA}

In this section, we describe Euclidean phylogenetic PCA \cite{revell2009size}. We define the tree structure with leaf node observations which constitutes the input to the algorithm. And we define the underlying Brownian motion model on the tree.

Phylogenetic PCA takes as input a rooted, bifurcating tree. We denote this by $$(V, E, L),$$ where $V$ is the node set, $E$ is the edge set, and \textit{L} is a function giving the 'length' of an edge, $L(e) = L(n_i, n_j) \in \mathbb{R}_{\geq 0}, e = (n_i, n_j) \in E$. A node $n \in V$  can have an associated value $x_n \in \mathbb{R}^d$. We will call such a tree a \textit{p-tree}. P-PCA assumes that only leaf nodes are observed. 
In this section, the node values will be Euclidean, so $x_i \in \mathbb{R}^d$. In the subsequent sections, we will generalize this to allow values in a Riemannian manifold.

The length of a path between two nodes is the sum of the edge lengths between nodes in the path. There is a unique shortest path from the root \textit{r} to any node $n_i \in V$, and we denote the length of this path by $L(r, n_i)$.

The \textit{most recent common ancestor} of two nodes $n_i$ and  $n_j$, $\text{MRCA}(n_i, n_j)$, is the unique node $n \in V$ from which there is a path to both $n_i$ and $n_j$, and which minimizes the path-lengths $L(n, n_i)$ and $L(n, n_j)$. 

We interpret a p-tree with leaf observations as a phylogenetic tree, where only the leaf nodes are recent enough in time to be known. % The nodes in the tree might represent [...]
Estimating phylogenetic trees (i.e. nodes, edges and edge-lengths) for various species is a developed and active field of research. See for example \cite{nyakatura2012updating}, containing the tree for the mammal jaws data set we analyse in section \ref{sect_applications}.

\subsection{The Brownian motion model on a tree}

In this section we describe the model of a p-tree with branches representing Brownian motions on $\mathbb{R}^d$, and how this leads to a joint normal distribution of the leaf nodes.

Let $\{B^{x_0}_t\}_{t \geq 0}$ denote a Brownian motion (BM) in $\mathbb{R}^d$ with initial value $x_0$. Given a p-tree $(V, E, L)$, the associated Brownian motion model assumes that each edge of the tree corresponds to the path of an $\mathbb{R}^d$-valued Brownian motion. All such Brownian motions are assumed to have the same covariance matrix. Namely, the values $x_{n_i}, x_{n_j} \in \mathbb{R}^d$ of two nodes $n_i, n_j \in V$ s.t. $(n_i, n_j) \in E$ and $L(n_i, n_j) = t'$ are related by

\begin{align}\label{BM_p_tree}
x_{n_j} \sim \mathbb{P}(B^{x_{n_i}}_{t'}),    
\end{align}
i.e. $x_{n_j}$ is an observation from the time-$t'$ transition distribution of $B^{x_{n_i}}_{t'}$. Thus all nodes except for the root are random variables. Figure 1b shows one realization of the p-tree in figure 1a.

The assumption that the nodes are endpoints of Brownian motions implies that the joint distribution of leaf-node values $\{x_1, \dots, x_N \}$ is a normal distribution on $\mathbb{R}^{N \cdot d}$. In \cite{felsenstein1973maximum}, it is shown that

\begin{align}
 (x_1, \dots, x_N) \sim N(D \hspace{0.3mm} r, R \otimes C),
\end{align}

\noindent a normal distribution with mean $Dr$ and covariance matrix $R \otimes C$. $D$ is an $(N\cdot d) \times d$ design matrix where $D_{ij} = 1$, if $(j-1)\cdot N < i \geq j \cdot N$, and $0$ otherwise. $r \in \mathbb{R}^d$ is the value of the root node.  $R \in \mathbb{R}^{d \times d}$ is the covariance matrix for each BM between nodes, with $R_{ij} = \sigma_{ij}$ the covariance between coordinate $i$ and $j$. $C \in \mathbb{R}^{N \times N}$ is the \textit{evolutionary covariance matrix} determined by the p-tree; $C_{ij} = L(r, \text{MRCA}(n_i, n_j))$. I.e. $C_{ij}$ is the shared branch length between $n_i$ and $n_j$. $\otimes$ is the Kronecker product of matrices.

The maximum likelihood estimates of the phylogenetic mean-parameter $r$ and the covariance parameter \textit{R} are,
\begin{align}\label{root_est}
 \hat{r} = \left(1^T C^{-1} 1 \right)^{-1} \left(1^T C^{-1} X \right) \in \mathbb{R}^d,  
\end{align}
\begin{align}\label{cov_est}
 \hat{R} = \frac{1}{N-1} \left(X -  \hat{r}^T\right)^T C^{-1} \left(X -  \hat{r}^T \right) \in \mathbb{R}^{d \times d},  
\end{align}
\noindent where $X \in \mathbb{R}^{N \times d}$ is the matrix where each row is a leaf-node value, and $X -  \hat{r}^T \in \mathbb{R}^{N \times d}$ is the row-wise difference  (see \cite{harmon2019phylogenetic}). Note that $\hat{R}$ is the ordinary empirical covariance matrix, but with the data centered around the phylogenetic mean $\hat{r}$ and weighted by $C$.

\subsection{Phylogenetic PCA}

Under the Brownian motion model on the p-tree, the leaf node observations $\{x_1, \dots, x_N\}$ are not independent. Their covariance is described by the matrix $C \otimes R$, which is block-diagonal (implying independence) only for a tree with two leaf nodes and no inner nodes. Given a p-tree for the data, we already know the matrix \textit{C}, i.e. we already know the part of the covariance that is caused by the tree-structure. Therefore, we are interested, rather, in estimating the covariance of the Brownian motions, $R$. Using this covariance matrix, one can proceed as in ordinary PCA and find the directions of highest variability.

We can now state the algorithm for phylogenetic PCA. In summary, it consists in doing ordinary PCA but with the data centered at the root estimate $\hat{r}$ and using $\hat{R}$ instead of the ordinary empirical covariance matrix. The algorithm below returns the principal component representation of the data in the original space $\mathbb{R}^d$. One might be more interested in a representation of the data directly in the lower dimensional space, i.e. a plot of the data in $\mathbb{R}^k$, $k < d$, wrt. a \textit{k}-dimensional basis of eigenvectors. This other use case is just a simple reformulation of the last step in the algorithm below. Furthermore, it is sensible  to choose \textit{k} based on a plot of the eigenvalues.

\vspace{4mm}
\begin{algorithm}[H]\label{p_PCA_algo}
  \SetAlgoLined
  \KwIn{A p-tree $(V, E, L)$ with leaf observations $\{x_i\}_{i=1..N}$, and a choice $k \in \{1, \dots, d\}$ of dimension of the reduced representation}
  \KwOut{A dimension-reduced representation of the data, $\{\tilde{x}_i\}_{i=1..N} \in M$.}
  
    \vspace{2mm}Compute ML estimates of the root $\hat{r}$ (eq. (\ref{root_est})) and center the data around $\hat{r}$, $$(x_i)_{\hat{r}} := x_i - \hat{r}, \quad i=1..N.$$
    
    \vspace{2mm} Compute the covariance estimate $\hat{R}$ (eq. (\ref{cov_est})).\vspace{2mm}
    
    Eigen-decompose $\hat{R}$ and proceed as in ordinary PCA: project the $\hat{r}$-centered observations $\{(x_i)_{\hat{r}}\}_{i=1..N}$ to the subspace spanned by the first \textit{k} eigenvectors, to get principal components $\{\tilde{x}_i\}_{i = 1..N}$.  \vspace{2mm}
    
    \Return $\{\tilde{x}_i\}_{i = 1..N}$
  \caption{Phylogenetic PCA}
\end{algorithm}
\vspace{4mm}

% Note that the eigenvalues in p-PCA depends on the edge-length unit, and it is not equal to the projected variance as in ordinary PCA. But the relative sizes of eigenvalues $j$ and $k$ still indicate how much more variance is explained by the $j$-dimensional eigenspace compared to the $k$-dimensional. [... slet evt. denne paragraf]

\section{Riemannian geometry}\label{sect_riemannian_geometry}

In this section, we give a brief introduction to Riemannian geometry, enough to provide an intuitive understanding of tangent p-PCA. For a more thorough introduction to Riemannian geometry, see e.g. \cite{lee2018introduction}. We include Riemannian Brownian motions and statistical methods on Riemannian manifolds.

A Riemannian manifold is a pair of objects, $(M, g)$, where $M$ is a smooth manifold and $g$ is a Riemannian metric. A smooth manifold is the generalization of a smooth surface in $\mathbb{R}^3$ to higher dimensions, i.e. an \textit{n}-dimensional manifold can be thought of as a smooth $n$-dimensional surface in $\mathbb{R}^d$, for some $d > n$. The word \textit{smooth}, which we will mostly omit henceforth, means infinitely differentiable. 
Locally around any point $p \in M$, the manifold can be identified with $\mathbb{R}^n$, where $n = \text{dim}(M)$. This local vector space representation of \textit{M} at \textit{p} is called the \emph{tangent space} of \textit{M} at $p$, and is denoted $T_p M$. $T_p M$ is the \textit{n}-dimensional plane best approximating the surface \textit{M} at \textit{p}. Another viewpoint is that $T_p M$ contains any possible tangent vector of smooth curves on \textit{M} passing through \textit{p}. Note that, taking the viewpoint of \textit{M} as embedded in $\mathbb{R}^d$, tangent vectors in $T_p M$ are elements of $\mathbb{R}^d$.

On a general smooth manifold, there is no notion of distance between points or of lengths of curves on \textit{M}. The length of a curve can be computed by measuring the length of its tangent vectors, which a \textit{Riemannian metric} makes possible. A Riemannian metric \textit{g} is a collection of inner products on the tangent spaces of \textit{M}, $g_p(\cdot, \cdot) : T_p M \times T_p M \to \mathbb{R}$, smooth as a function of \textit{p}. The length of a tangent vector can now be computed as the norm induced by this inner product, $\Vert \dot{\gamma}(t_0) \Vert_g := \sqrt{g_{\gamma(t)}(\dot{\gamma}(t), \dot{\gamma}(t))}$, where $\gamma$ is a smooth curve on \textit{M} and $\dot{\gamma}(t)$ is its tangent vector at time \textit{t}. The length of a curve $\gamma: [0,1] \to M$ can be computed as $L(\gamma) = \int_0^1 \Vert \dot{\gamma}(t) \Vert_g dt$. From the notion of curve length we can define a distance metric on \textit{M}, 
\begin{align*}
    d_g(p, q) = \{L(\gamma) \hspace{1mm} \vert \hspace{1mm} \gamma \text{ is the shortest curve s.t. } \gamma(0)=p, \gamma(1)=q\}, \quad p, q \in M.
\end{align*}

\noindent Note that for $M = \mathbb{R}^n$ and \textit{g} the Euclidean metric (i.e. the metric s.t. $g_p$ is the standard Euclidean inner product for all $p \in \mathbb{R}^d$) this reduces to $d_g(p,q) = \sqrt{(p - q)^T(p - q)}$, the ordinary Euclidean distance between $p$ and $q$.

\subsection{Geodesics, $\exp$ and $\log$}

A central notion in Riemannian geometry is that of a \textit{geodesic}, which is a locally length-minimizing curve. This means that if $\gamma$ is a geodesic, any sufficiently small (smooth) perturbation of $\gamma$ will increase $L(\gamma)$. 

A geodesic solves a second order ODE, and is uniquely determined by its initial position and tangent vector. Thus we can write $\gamma_p^v(t)$ for a geodesic $\gamma$ starting at $p \in M$ with initial tangent $v \in T_p M$. For fixed time $t=1$ we can consider $\gamma_p^v(1)$ to be a map from the tangent space $T_p M$ to $M$, called the Riemannian \textit{exponential map},
\begin{align}
    \exp_p : T_p M \to M : v \mapsto \exp_p(v) := \gamma_p^v(1).
\end{align}

At any $p\in M$, there exists an open subset $V \subset T_p M$ such that the Riemannian exponential is a diffeomorphism onto $\exp_p(V) := U \subset M$. The largest such subset \textit{V} is called the \textit{tangential cut locus} at \textit{p}, denoted $\mathdutchcal{C}_p$. $\exp_p$ has a smooth inverse on $C_p := \exp_p(\mathdutchcal {C}_p)$, called the \textit{cut locus},
\begin{align}
    \log_p : C_p\subset M \to T_p M : q \mapsto \log_p(q) := \exp^{-1}_p(q).
\end{align}
\noindent Thus $\log_p(q)$ is the tangent vector satisfying $\gamma_p^{\log_p(q)}(1) = q$. On the sphere (with the canonical pull-back metric), $C_p = S^2 \setminus \{-p\}$, i.e. all points except the antipodal point. On a general Riemannian manifold we do not have an explicit characterization of $\mathdutchcal{C}_p$ and $C_p$.

\subsection{Brownian motions on a Riemannian manifold}\label{sect_riem_BM}

A Riemannian Brownian motion is a stochastic process on $(M, g)$ whose infinitesimal generator is the Laplace-Beltrami operator \cite{hsu2002stochastic}. To simulate a Brownian motion via a Euler-Maruyama-like scheme \cite{said2012brownian}, choose some initial point $p_0 \in M$ and a step-size $\Delta$. We will denote the generated Brownian motion by $\left\{B^{p_0}(t) \right\}_{t \geq 0}$. Choose an orthonormal basis of $T_{p_0}M$ wrt. the Riemannian metric. Draw a vector $v \in T_{p_0}M$ from a standard normal distribution wrt. this basis. Generate the next point as $B^{p_0}(\Delta) = \exp_{p_0}\left(\frac{v}{\Vert v \Vert_g} \Delta \right) \in M$. At any time-point \textit{t} of the trajectory, the next point $B^{p_0}(t + \Delta)$ is generated as above by letting the initial point be $B^{p_0}(t)$. The resulting process will be an approximation of a Riemannian Brownian motion. 

A non-isotropic Brownian motion (i.e. with non-identity covariance) can be generated similarly, but the construction is more involved since the orthonormal basis cannot be chosen arbitrarily in each tangent-space, it needs to be consistently transported between them. We refer to \cite{pennec2019riemannian}, chapter 10.

\subsection{Riemannian geometric statistics: the Fréchet mean and tangent PCA}\label{sect_riem_geom_stats}

The field of \textit{geometric statistics} deals with observations lying on a manifold - often assumed to be a Riemannian manifold. Given observations $x_1, \dots, x_N$ on a Riemannian manifold, we can define a mean point $\mu \in M$ as a minimizer of the sum of squared geodesic distances to the observations,
\begin{align}\label{frechetMean}
    \mu = \underset{\bar{\mu} \in M}{\text{argmin}} \sum_{i=1}^N d_g(\bar{\mu}, x_i)^2.
\end{align}
This is called the \textit{Fréchet mean} \cite{pennec2019riemannian}. Note that this might not be unique, for uniqueness conditions see \cite{pennec2019riemannian}, \cite{kendall1990probability}.

A general strategy for applying a Euclidean statistical method to observations on a manifold is to first map the observations to the tangent space of the mean, $T_\mu M$, via the Riemannian logarithm. Then apply the Euclidean method in $T_\mu M$, and map the result back to \textit{M} via the Riemannian exponential. \textit{Tangent PCA} (\cite{pennec2019riemannian}, \cite{fletcher2004principal}) does exactly this. Given some basepoint $\mu$, e.g. the Fréchet mean, the procedure consists in doing ordinary PCA in $T_{\mu} M$. The observations in $T_{\mu} M$ can thus e.g. be projected to the desired \textit{k}-dimensional PCA subspace and mapped back to \textit{M} via the exponential map. 

Tangent phylogenetic PCA will be defined in an analogous way. First, we iteratively estimate the root node $r$, using the Euclidean estimator in each tangent space. Then we apply Euclidean p-PCA in the tangent space $T_{r}M$, and map the result back to $M$. 

\section{Tangent phylogenetic PCA}\label{sect_tangent_p_PCA}

We now assume given a p-tree $(V, E, L)$ where the leaf observations $\{x_i\}_{i = 1 .. N} \in M$ are points on a Riemannian manifold $(M, g)$.

The model of Euclidean Brownian motions structured according to a p-tree is directly transferable to the manifold setting. 
Thus we assume the data $\{x_i\}_{i=1..N}$ to be observations of leaf nodes on $M$.

As for Euclidean p-PCA, the first step of tangent p-PCA is to estimate the root $r\in M$. Algorithm 2 describes a method based on iterative tangent space approximations of \textit{M}. It is inspired by a gradient descent algorithm for estimating the Fréchet mean (\cite{pennec2019riemannian}, section 2.2.2, and \cite{pennec2006riemannian}). 

\begin{algorithm}
  \SetAlgoLined
  \KwIn{A p-tree $(V, E, L)$ with leaf observations $\{x_i\}_{i=1..N}$, an initial guess $r_0 \in M$ and a convergence threshold $\epsilon > 0$.}
  \KwOut{An estimate of the root $\hat{r} \in M$.}
  
    \vspace{2mm} Set $\hat{r} := r$ and choose some initial $\tilde{r} \in T_{r}M$ s.t. $\Vert \tilde{r} \Vert_g > \epsilon$.
    
    \vspace{2mm}

    \While{$\Vert \tilde{r}\Vert_g > \epsilon$}{

    \vspace{2mm}Map the data to $T_{\hat{r}} M$, $({x}_i)_{\hat{r}} := \log_{\hat{r}}(x_i)$, $i=1..N$, and collect these vectors in a matrix $X \in \mathbb{R}^{N \times d}$.
    
    \vspace{2mm}Compute $\tilde{r} = \left(1^T C^{-1} 1 \right)^{-1} \left(1^T C^{-1} X \right) \in T_{\hat{r}}M,$ the Euclidean estimate from eq. $(\ref{root_est})$.
    
    \vspace{2mm}Compute $\hat{r} = \exp_{\hat{r}}(\tilde{r}) \in M$\vspace{2mm}
    }
{ \Return $\hat{r}$
}
  \caption{Root node estimation}
\end{algorithm}

The second step of tangent p-PCA is to map the data to the tangent space of the estimated root node via the Riemannian logarithm, and then do Euclidean p-PCA. However, there is a caveat to this: we need to compute the covariance matrix using a canonical basis, and we need the eigenvectors to be orthonormal wrt. the Riemannian metric \textit{g}. This can be ensured by representing the observations, $({x}_i)_{\hat{r}} := \log_{\hat{r}}(x_i) \in T_{\hat{r}}M, i=1\dots N,$ wrt. a basis that is orthonormal wrt. the Riemannian metric -  the result will not depend on which orthonormal basis is used. Let $g_{\hat{r}}$ be the $d \times d$ matrix representing the Riemannian metric locally around $\hat{r}$. Then an orthonormal basis for $T_{\hat{r}}M$ can be found via the Cholesky decomposition $g^{-1}_{\hat{r}} = LL^T$. The columns of \textit{L} is an orthonormal basis wrt. \textit{g}. Tangent p-PCA is described in algorithm 3 below.

\begin{algorithm}
\caption{Tangent phylogenetic PCA}\label{alg:two}
\KwIn{A p-tree $(V, E, L)$ with leaf observations $\{x_i\}_{i=1..N}$, an estimate of the root node $\hat{r}$ and a dimension $k \in \{1, \dots, d\}$ of the reduced representation.}
\KwOut{A \textit{k}-dimensional  representation of the data, $\{\tilde{x}_i\}_{i=1..N} \in M$.}
\vspace{2mm}Map the data to $T_{\hat{r}} M$, $({x}_i)_{\hat{r}} := \log_{\hat{r}}(x_i)$, and collect these vectors in a matrix $X \in \mathbb{R}^{N \times d}$.

\vspace{2mm}Compute the Cholesky decomposition \textit{L} of $g^{-1}_{\hat{r}} $, $g^{-1}_{\hat{r}} = L L^T$, and change coordinates of the observations to the Cholesky basis, i.e. $X_{ortho} := X(L^{-1})^T.$

\vspace{2mm}Compute $\hat{R} = \frac{1}{N-1} X_{ortho}^T C^{-1} X_{ortho} $, the phylogenetic covariance matrix.

\vspace{2mm}Eigen-decompose $\hat{R}$ and proceed as in ordinary PCA in $T_{\hat{r}}M$: project the observations $X_{ortho}$ to the span $\Sigma^k \subset T_{\hat{r}}M$ of the first \textit{k} eigenvectors to get principal components $\bar{X}_{ortho} $ in $T_{\hat{r}}M$. \vspace{2mm}

\vspace{2mm} Write the principal components wrt. to Euclidean coordinates, i.e. $\bar{X} := XL^T$. Row $i$ of $\bar{X}$ is principal component $\bar{x}_i \in T_{\hat{r}}M$.\vspace{2mm}

 \Return $\{\tilde{x}_i\}_{i = 1..N} := \{\exp_{\hat{r}}(\bar{x}_i)\}_{i = 1..N}$.
\end{algorithm}

Note that, for $M = \mathbb{R}^d$ and the Euclidean metric $g$, algorithms 2 and 3 yields the Euclidean root estimates and p-PCA, respectively.

If $V \subset T_p M$ is a \textit{k}-dimensional subspace , $\exp_p(V \cap C_p) \subset M$ is a \textit{k} dimensional submanifold. If $\{(x_i)_{\hat{r}}\}_{i=1..N} \subset C_{\hat{r}}$, then $\{\tilde{x}_i\}_{i = 1..N} \subset M$ are points on the submanifold $\exp_{\hat{r}}(\Sigma^k \cap C_{\hat{r}})$. In this sense they constitute a \textit{k}-dimensional representation of the original data.

\section{Simulations and applications}\label{sect_applications}

In this section, we investigate the behaviour of tangent p-PCA via simulations on the sphere and by applying it to a data set of mammal jaws, represented as landmark shapes. All computations are done using Jaxgeometry  https://bitbucket.org/stefansommer/jaxgeometry/. The implementations are based on automatic differentiation libraries, as is further described \cite{kuhnel2019differential}.

\subsection{Simulations on the sphere}

We generate observations on the unit sphere, $S^2$, by simulating isotropic Brownian motions on $S^2$ structured according to a p-tree. The topology of the tree is equivalent to the one on figure 1a, and branch lengths are proportional. We simulate isotropic Brownian motions using the Euler-Maruyama-like scheme described in section \ref{sect_riem_BM}. The true root is the north pole, and the convergence criterion parameter is set to $\epsilon = 10^{-5}$. For each of the 1000 simulated trees, we estimate the root node via algorithm 2 and compute its Riemannian distance to the true root. 
 A single realization of a p-tree is shown in figure 3. The error-distribution is shown in figure 2.
 We note that it resembles a $\chi^2$-distribution. When the leaf nodes are Euclidean, the root is the mean of a Normal distribution, and it is well known that the exact error-distribution is chi-square.

\begin{figure}[h!]
\centering
\begin{minipage}[t]{.5\textwidth}
  \centering
  \includegraphics[width=1\linewidth]{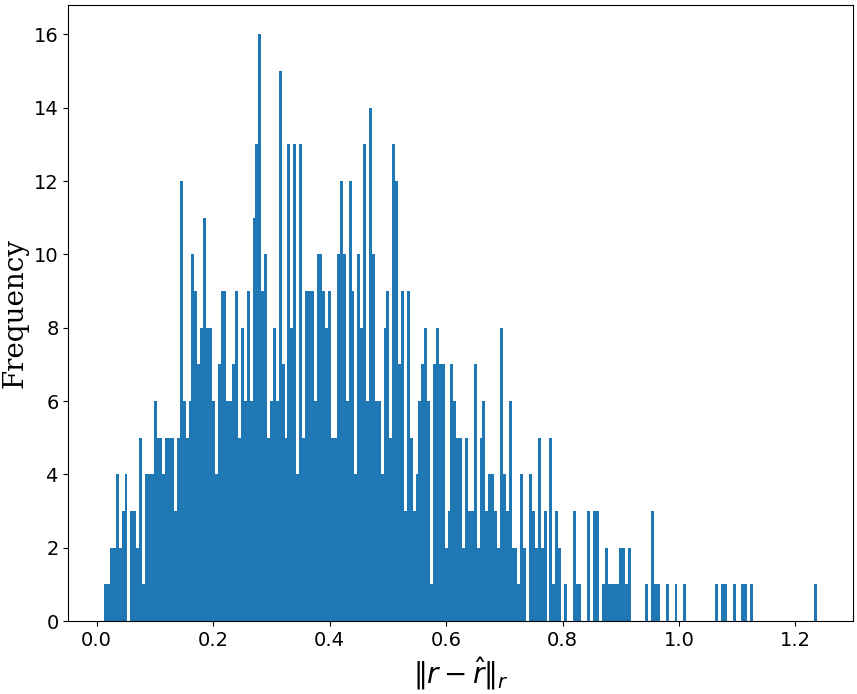}
  \captionof{figure}{Distribution of Riemannian distance $\Vert r - \hat{r}\Vert_r$, i.e. the geodesic distance between the true root and the root estimate, based on 1000 simulated trees on the sphere. For reference, the geodesic distance from the north pole (the true root) to a point on the equator is $\pi/2 \approx 1.57$.}
  \label{fig:test1}
\end{minipage}%
\begin{minipage}[t]{.5\textwidth}
  \centering
  \includegraphics[width=0.8\linewidth]{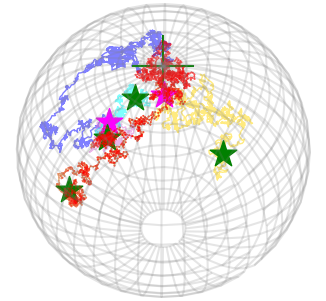}
  \captionsetup{width=.9\linewidth}
  \captionof{figure}{A realization of the tree in figure 1a, generated from Brownian motions on the sphere. The green '+' is the root node. The two pink stars are the inner nodes. The 4 green stars are the leaf nodes.}
  \label{fig:test2}
\end{minipage}
\end{figure}

 Figure 4 shows a histogram of the number of iterations until convergence for two different choices of initial root-value: the Fréchet mean and the south pole, respectively. 

\subsection{Mammal jaws data set}\label{sect_mammal_jaws}

In this section, we analyse a data set of mammal jaws. For each of 113 mammal species, the data set contains from 1 to 8 jaws, each represented by 14 landmarks in $\mathbb{R}^2$. Figure $\ref{mammal_jaws_roots}$ shows the 113 landmark shapes represented in $\mathbb{R}^2$. For a full description of the data, see \cite{conith2018influence}. The phylogenetic tree is estimated in \cite{nyakatura2012updating}.

The first step of our analysis is to perform Procrustes alignment of the shapes. After that, we compute the Euclidean mean of the jaw observations within each species, and proceed with these 113 points instead of the full data set. After that, we consider them as points in the LDDMM landmark space with a Gaussian kernel $k(q_i, q_j) := \beta \exp^{- \Vert q_i - q_j \Vert^2 / 2 \sigma^2}\in \mathbb{R}$, $q_i \in \mathbb{R}^2$ (see \cite{pennec2019riemannian}, \cite{younes2010shapes}, \cite{miller2002metrics}). This a Riemannian manifold $(M, g)$ where $M = \mathbb{R}^{2\cdot 14} = \mathbb{R}^{28}$ and the metric is given by $g_p(u, v) = u^T K_p^{-1} v$. Here, $K_p$ (the so-called co-metric) is a block matrix $K_p = \left[(\tilde{K}_p)_{i,j}\right]_{i,j = 1..14}$, with 
\begin{align*}
    (\tilde{K}_p)_{ij} = 
    \begin{bmatrix} 
    k(q_i, q_i) q_i^T q_i & k(q_i, q_j) q_i^T q_j \\
    k(q_j, q_i) q_j^T q_i & k(q_j, q_j) q_j^T q_j,
    \end{bmatrix}
\end{align*}

\noindent for shape $p = (q_1, \dots, q_{14})$ consisting of 14 landmarks $q_i \in \mathbb{R}^2$.

Here, the parameter $\sigma$ determines the width of the kernel; if $\sigma$ is large, landmarks far apart will be affected by perturbing one of them. In our analysis, we set $\beta = 1$, and $\sigma$ equal to 1.5 times the average over all shapes of the average Euclidean distance between landmarks in each shape. I.e. $\sigma = 0.208$ is proportional to the typical distance between landmarks within a typical jaw shape.

We start out by estimating the root node using algorithm 2. The results are presented in figure \ref{mammal_jaws_roots}. As initial estimate we use the root estimate from Euclidean p-PCA. The algorithm converges in 8 iterations with a threshold of $\epsilon = 10^{-4}$. The root estimate deviates visibly from the Euclidean root estimate.

We then perform tangent p-PCA according to Algorithm 3. 
We compute the phylogenetic covariance matrix $\hat{R}$ in $T_{\hat{r}}M$ and plot its eigenvalues to determine a dimension \textit{k} which describes most of the variability, while reducing the dimension significantly. The eigenvalues are plotted in figure 5. The first 6 eigenvectors describes a large proportion of the (non-phylogenetic) variation, thus we choose to represent the observations in $k=6$ dimensions. Figure \ref{PCA_comparison_mammals} below shows the projected observations for both tangent p-PCA and for Euclidean p-PCA on Procrustes aligned landmark shapes.

\begin{figure}[H]
\centering
\begin{minipage}[t]{.5\textwidth}
  \centering
  \includegraphics[width=0.9\linewidth]{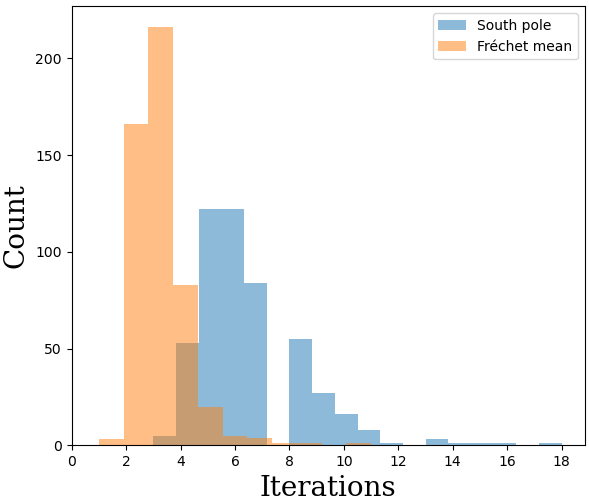}
  \captionof{figure}{Histogram showing the number of iterations until convergence of the root-estimate on the sphere, for two different choices of initial value: the Fréchet mean (yellow) and respectively the south pole (blue). For each initial value, 500 trees was simulated.}
  \label{fig:test1}
\end{minipage}%
\begin{minipage}[t]{.5\textwidth}
  \centering
  \includegraphics[width=1\linewidth]{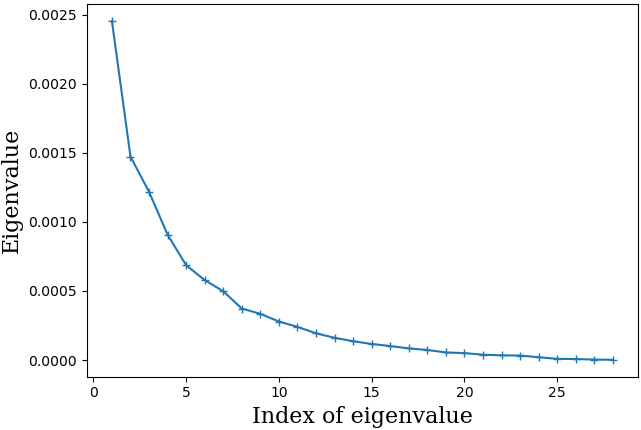}
  \captionsetup{width=.9\linewidth}
  \captionof{figure}{Eigenvalues from tangent p-PCA on the Mammal jaws data set.}
  \label{fig:test2}
\end{minipage}
\end{figure}

\begin{figure}[H]
\centering
\includegraphics[width=1\textwidth]{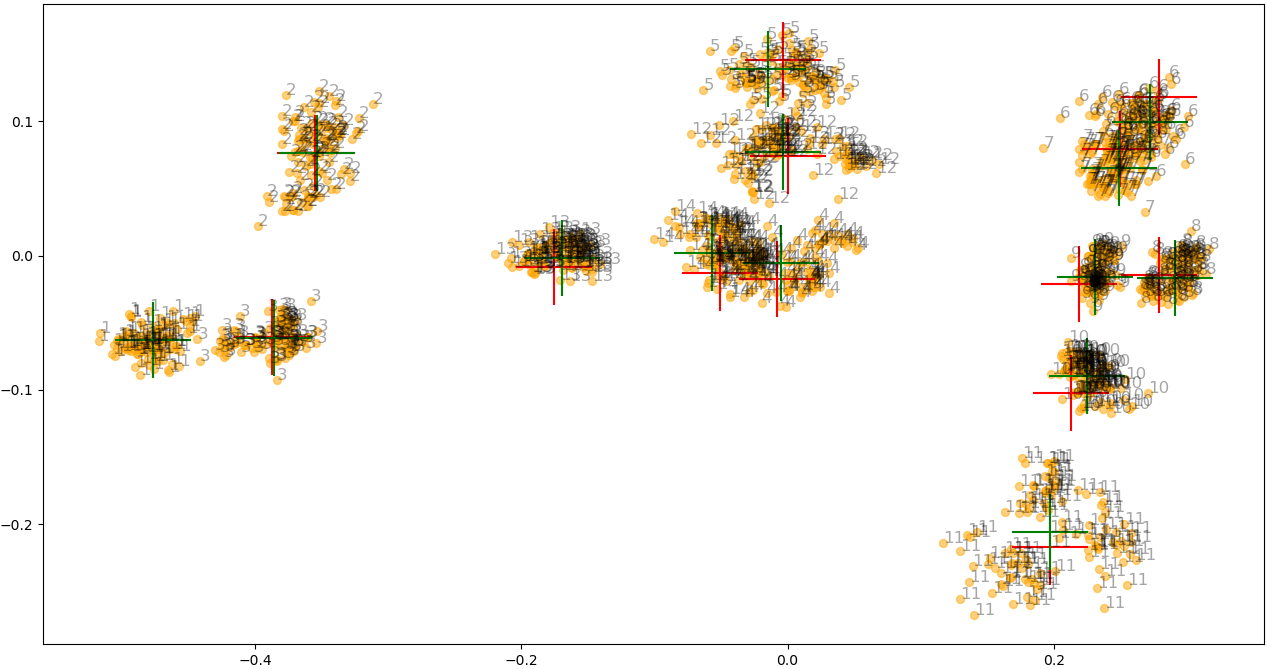}
\caption{Illustration of the Mammal jaws data set (yellow dots) and 2 root estimates, one by algorithm 2 (green '+'-signs) and one by Euclidean p-PCA (red '+'-signs) after Procrustes alignment. Each shape consists of 14 landmarks (dots), numbered in the figure.\vspace{10mm}}
\label{mammal_jaws_roots}
\includegraphics[width=1\textwidth]{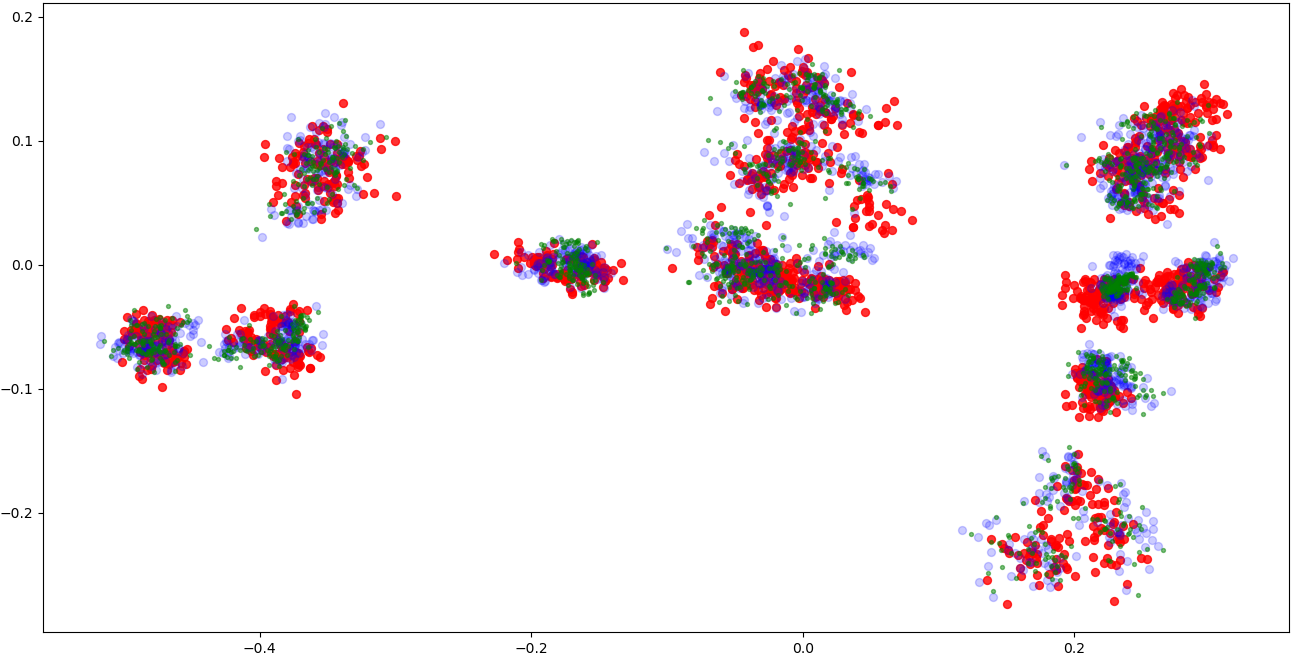}
\caption{Blue dots are the Mammal jaws observations after Procrustes alignment. Red dots are the observations in $T_{\hat{r}}M$ projected to the first 6 tangent p-PCA eigenvectors and then mapped back to \textit{M} via the Riemannian exponential. Green dots are the observations projected to the first 6 Eigenvectors of Euclidean p-PCA.}
\label{PCA_comparison_mammals}
\end{figure}

\section{Conclusions and future work}\label{sect_discussion}

We have constructed a version of phylogenetic PCA for manifold-valued data and applied it to a dataset of landmark shapes. We have argued that there is a need in evolutionary biology for doing exactly this. The method is based on Euclidean approximations in a way which respects the manifold structure of the data while keeping the interpretation of Euclidean p-PCA.

Computationally, the most demanding part of tangent p-PCA is the root-estimation, specifically the computation of logarithms.  In each iteration, \textit{N} Riemannian logarithms must be computed, each of which is an optimization problem on $\mathbb{R}^{\text{dim}(M)}$. However, the log-computations are independent of each other, so can be done in parallel.

\bibliographystyle{plain}
\bibliography{refs.bib}

\end{document}